# GROEBNER BASES IN JAVA
# WITH APPLICATIONS IN COMPUTER GRAPHICS

**Branko J. Malešević[1], Ivana V. Jovović[2], Milan Z. Čampara[3]**

**Abstract**

In this paper we present a Java implementation of the algorithm that computes Buchbereger's and reduced Groebner's basis step by step. The Java application enables graphical representation of the intersection of two surfaces in 3-dimensional space and determines conditions of existence and planarity of the intersection.

## 1. GROEBNER BASES - RETROSPECTIVE

We will focus on 3-dimensional Euclidean space. The main topic of this paper is planarity of the intersection of a system of polynomial equations

(1) $\quad f_1(x, y, z) = 0, \ldots, f_m(x, y, z) = 0$;

where $f_1, \ldots, f_m$ are polynomials in three variables over the real field $\mathbb{R}$. We use the Groebner bases technique as a follow-up of the papers [1], [2]. The basic notations of this technique are established in [1-6].

[1]Branko Malešević, Ph.D., docent, Faculty of Eletrical Engineering, University of Belgrade, E-mail: malesevic@etf.bg.ac.rs
[2] Ivana Jovović, teaching assistant, Faculty of Eletrical Engineering, University of Belgrade, E-mail: ivana@etf.bg.ac.rs
[3]Milan Čampara, senior student at the Faculty of Eletrical Engineering, University of Belgrade, E-mail: milan_campara@yahoo.com

We are interested to know whether system (1) has an intersection in the terms of Groebner Bases. The answer is: system (1) is consistent iff for the reduced Groebner Basis (RGB) holds :

(2) $$RGB \neq \langle 1 \rangle.$$

We will consider only systems (1) with nonempty intersection.

A starting $m$-tuple of polynomials $F = (f_1, \ldots, f_m)$ determines the ideal (see [6]):

(3) $$I = \left\{ \sum_{i=1}^{m} h_i f_i : h_1, \ldots, h_m \in R[x, y, z] \right\},$$

that is denoted by $\langle f_1, \ldots, f_m \rangle$. The polynomials $f_1, \ldots, f_m$ are called generators for the ideal $I = \langle f_1, \ldots, f_m \rangle$. Let us review Hilbert basis theorem. It claims that every ideal has a finite generating set (see [6]). Furthermore, for the ideal $I = \langle f_1, \ldots, f_m \rangle$ we define the ideal of leading terms:

(4) $$\langle LT(I) \rangle = \langle \{LT(f) \mid f \in I\} \rangle,$$

where $LT$ denotes the leading term in the monomial order (see [3], [4]). We will restrict our discussion to the lexicographical order $x > y > z$ (see [2], [3]).

For any ideal $I = \langle f_1, \ldots, f_m \rangle$ it is true:

(5) $$\langle \{LT(f_1), \ldots, LT(f_m)\} \rangle \subseteq \langle LT(I) \rangle.$$

A finite generating set $G = \{g_1, \ldots, g_k\}$ of the ideal $I = \langle f_1, \ldots, f_m \rangle$ is called a *Groebner basis* (GB) for $I$ if:

(6) $$\langle \{LT(g_1), \ldots, LT(g_k)\} \rangle = \langle LT(I) \rangle.$$

In his dissertation, B. Buchberger has proved that for each ideal there exists a GB.

We present the algorithm in pseudo code (see [2], [6]):

Input: $F = \{f_1, \ldots, f_m\}$ - Set of generators of $I$
Output: $G = \{g_1, \ldots, g_k\}$ - GB for $I$ ($G \supseteq F$)

$G := F$
repeat
 $G' := G$
(*) for each pair $\{p, q\}$, $p \neq q$, in $G'$ do
 $S := S(p, q)$ - Syzygy of p and q
 $h := REM(S, G')$ - Remainder after dividing S with G'
 if $h \neq 0$ then $G := G \cup \{h\}$
until $G = G'$

For different sets of generators of the ideal $I$, the GB calculated by Buchberger's algorithm is not unique. GB is minimal iff $LC(g) = 1$ for all $g \in G$, where $\{LT(g) \mid g \in G\}$ is a minimal basis of the monomial ideal $\langle \{LT(f) \mid f \in I\} \rangle$ (see [2]). The *minimal Groebner basis* ($MGB$) is not unique as well. The r*educed Groebner basis* ($RGB$) is a $MGB$ for the ideal $I$ such that for all $p \in G$ there is no monomial of $p$ that belongs in $\langle \{LT(g) \mid g \in G \setminus \{p\}\} \rangle$ (see [5], [6]). The procedure for finding the $MGB$ is presented below.

Let $LC(g) = 1$ for all $g \in G$. If some $p \in G$ has the property $LT(p) \in \langle \{LT(g) \mid g \in G \setminus \{p\}\} \rangle$, then remove $p$ from $G$ (see [5]).

The reduction algorithm we apply to the $MGB$ and it is based on:

(**) If some other (non-leading) monomial term in $p$ is in $\langle \{LT(g) \mid g \in G \setminus \{p\}\} \rangle$, then replace $p$ from $G$ with $h_p$ - partial reimander[#] of $p$ by $G \setminus \{p\}$.

Reduction is completed by repeating (**) and as a result we obtain $RGB$ that is uniquely determined (see [5], [6]).

---

[#] related to the monomial term from $p$

## 2. AN ALGORITHM FOR PLANARITY OF INTERSECTION OF SURFACES

Let us consider the consistent system of two real non-linear polynomial equations in three variables:

(7) $$f_1(x, y, z) = 0 \wedge f_2(x, y, z) = 0.$$

System (7) has a *planar solution* if there is a linear-polynomial

(8) $$g = g(x, y, z) = Ax + By + Cz + D,$$

such that every solution of system (7) is also a solution of the linear equation

(9) $$g(x, y, z) = 0,$$

for some real constants $A, B, C, D$.

Thus we indicate that the system has a *planar intersection*. In accordance with paper [4], for the lexicographic order $x > y > z$, we have:

**Theorem 2.1.** *If a GB of system* (7) *contains linear polynomial* (8), *then the solution of system* (7) *is a planar.*

**Theorem 2.2.** *Let system* (7) *have the planar intersection*

(10) $$Ax + By + Cz + D = 0,$$

*for some* $A, B, C, D \in \mathbb{R}$ *and* $A \neq 0$. *If for the ideal* $I = \langle f_1, f_2 \rangle$ *is true*

(11) $$y \notin \langle LT(I) \rangle \wedge z \notin \langle LT(I) \rangle,$$

*then the linear polynomial*

(12) $$\hat{g} = \hat{g}(x, y, z) = x + (B/A)y + (C/A)z + (D/A)$$

*is an element of the* RGB.

On the count of Buchberger's and the reduction algorithm we can generate the statement:

**Algorithm of planarity.** *If for an ideal* $I = \langle f_1, f_2 \rangle$ *of system* (7) *exists a generator* $h$ *in* Buchberger's algorithm *which is linear or if by applying the reduction algorithm we obtain a reduced generator* $h_p$ *which is linear, then the system has a planar intersection.*

Exactness of the previous algorthm is based on Theorem 2.1. Theorem 2.2. gives us a criterion for the occurrence of a linear polynomial in the final list of generators of the $RGB$.

The latest CAS (Computer Algebra Systems) works only with the monomial ordering for computing the $RGB$ (see [6]). The result is obtained without showing the intermediate steps. We present a program that guides us through the Buchberger's and the reduction algorithm step by step made in the Java environment, which is able to isolate the linear generators. The illustration of the program will be given in the last paragraph. The next two examples demonstrate the algorithm of planarity.

The first example deals with the linear generator which occurs in Buchberger's algorithm and still remains in the final list of generators in the $RGB$ when we use the reduction algorithm.

**Example 1.** Let the starting generators be

$$f_1 = -4x^2 - 9y^2 + z, \, f_2 = 4x^2 + 9y^2 - 2x - 3y.$$

Applying Buchberger's algorithm we add two new generators

$$f_3 = 2x + 3y - z, \, f_4 = -18y^2 + 6yz - z^2 + z.$$

The linear generator $f_3$ leads us to the conclusion that the intersection is planar.

It provides: $GB = \{-4x^2 - 9y^2 + z, 4x^2 + 9y^2 - 2x - 3y,$
$2x + 3y - z, -18y^2 + 6yz - z^2 + z\}.$

The reduction algoritm yields us to the $RGB$:

$$RGB = \left\{ x + \frac{3}{2}y - \frac{1}{2}z,\ y^2 - \frac{1}{3}yz + \frac{1}{18}z^2 - \frac{1}{18}z \right\}.$$

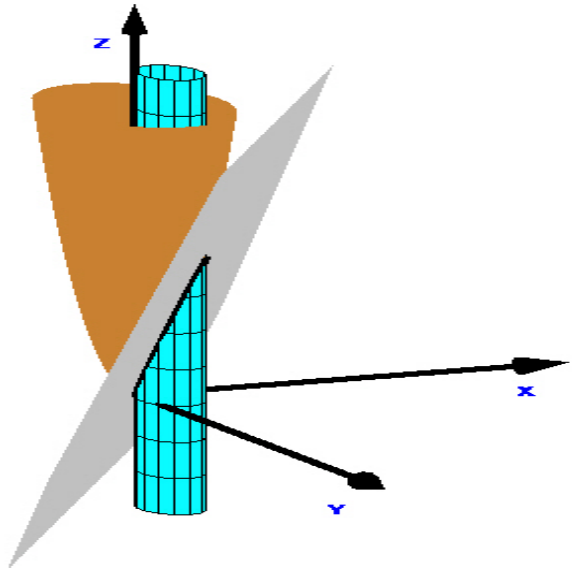

A graphic of the surfaces $f_1 = 0, f_2 = 0$ with the intersection plane.

The second example deals with the linear generators which occur neither in Buchberger's algorithm nor in the final step of the reduction algorithm. But yet, in the application of the reduction algoritm we can distinguish a linear polynomial.

**Example 2.** Let the starting generators be

$$f_1 = x + yz + y - z^4 - 4,\quad f_2 = y - z^3 - 1.$$

Following Buchberger's algorithm we obtain the Groebner basis:

$$GB = \{x + yz + y - z^4 - 4,\ y - z^3 - 1\},$$

which is also minimal. Applying the reduction algorithm to the

polynomial $p = x + yz + y - z^4 - 4$ we obtain the linear polynomial

$$h_p = x + y + z - 4$$

and we deduce that the intersection is planar.

Furthermore, if we assume that $p := h_p$, then we can reduce the polynomial $p = x + y + z - 4$ to the polynomial $h_p = x + z^3 + z - 3$ and the algorithm terminates. Therefore, the $RGB$ is:

$$RGB = \{x + z^3 + z - 3, y - z^3 - 1\}.$$

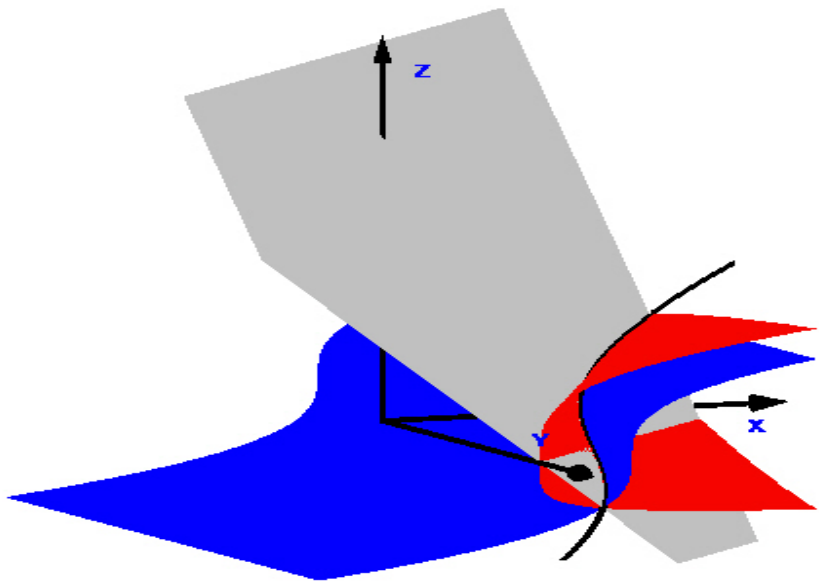

A graphic of the surfaces $f_1 = 0, f_2 = 0$ with the intersection plane.

It is worth pointing out that the Groebner bases technique has a wide range of applications related to the problems of planarity of the intersection of algebraic curves in Descriptive Geometry and Engineering Graphics (see [7, 8]).

# 3. GROEBNER BASES IN JAVA – STEP BY STEP

This application was developed for calculating the Groebner basis of an ideal and for representing it in 3D. The program was written entirely in Java. It can be run from any computer that has an up-to-date version of Sun Microsystems's free Java Virtual Machine (JVM). JVM enables a set of computer software programs and data structures to use a virtual machine model for the execution of other computer programs and scripts. The model used by a JVM accepts a form of computer intermediate language commonly referred to as Java bytecode.

The application itself is very easy to use. It consists of three text areas. In the left one, we insert the generators in lexicographical order, without any spaces separating the generators with the ENTER key. By pressing the "Calculate GB" button, depending on how long the calculation is, we can notice that the middle and right text areas are not empty anymore. The middle text area leads us through all of the steps of Buchberger's algorithm. As we know, this algorithm will generate a GB that usually isn't the minimal, nor the reduced GB. Because of the fact that they are very interesting and important for this field of study, we added a third text area on the right that calculates both the minimal and reduced GB.

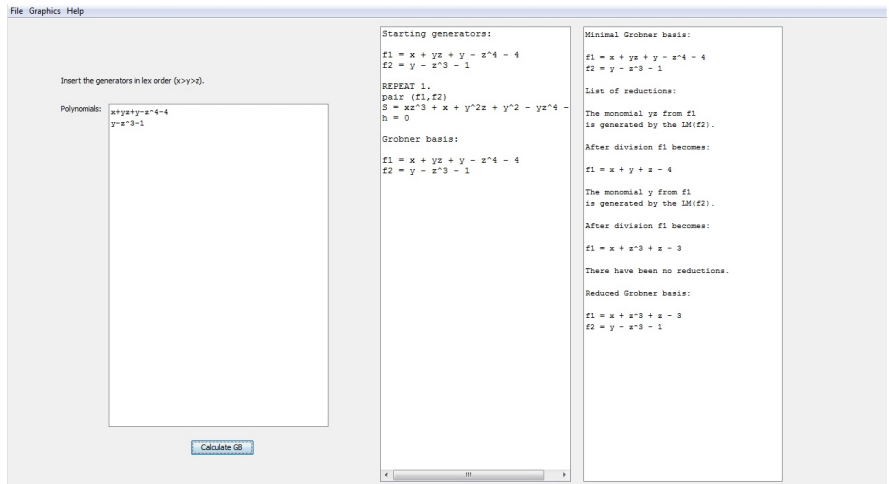

A screenshot of the application after calculation.

Now we can see the GB in 3D by selecting the "create file" item from the Graphics Menu. This will create three dpg files one for each phase of calculating (start GB, Buchberger's GB and reduced GB). These files can be viewed in a small but very functional program called DPGraph (http://www.dpgraph.com).

Let us examine how the program was organized. The main problem was how to make the computer understand what the user entered. First we created the monomial class that contains the coefficient and the power of x, y and z of the monomial. The next step was the polynomial class that has a list of monomials in lexicographical order. With this implementation, monomial and therefore polynomial operations are very simple. For instance, the division of two monomials is another monomial whose power of x is equal to the power of x of one monomial subtracted with the power of x of the second monomial. The same thing is for the power of y and z. The resulting monomial is zero if the monomials are not dividable. Now we created the Syzygy and Rem classes that are very vital for calculating the GB. Thanks to these classes, Buchberger's algorithm was implemented by translating the algorithm in pseudo code to Java source code. Once that was completed, we developed the algorithm for calculating the minimal and reduced GB. After running a few tests on the program, we concluded that it is more than necessary to see how this would look like in a 3D surrounding. After studying few 3D programs, we decided to use DPGraph and its tools for the 3D representation.

**Remark.** This Java application for Groebner bases will be available on the website http://symbolicalgebra.etf.bg.ac.rs/.

## ACKNOWLEDGEMENTS

The first and second author are supported by the project MNTRS, Grant No. ON144020.